\documentclass[12pt]{article}
\usepackage{graphicx}
\usepackage{cancel}[]
\usepackage{amssymb,amsmath}
\usepackage{ifpdf}

\usepackage{fullpage}
\usepackage{color}
\usepackage{ulem}
\usepackage{verbatim}
\usepackage{float}
\ifpdf
    \DeclareGraphicsExtensions{.pdf}
\else
    \DeclareGraphicsExtensions{.eps}
		
\fi

\usepackage{graphicx,epsfig}
\usepackage{cancel}[]
\usepackage{amssymb,amsmath}
\usepackage{subfigure}
\def\lsim{<\kern-2.5ex\lower0.85ex\hbox{$\approx$}\ }
\def\rsim{>\kern-2.5ex\lower0.85ex\hbox{$\approx$}\ }

\def\LAMBDABAR
{\hbox{$\lambda$\kern-0.52em\raise+0.45ex\hbox{--}\kern+0.2em}}

\vspace{.15in}
\begin{document}

\centerline{\large\bf{Proposed Search for a wind of Axion-like-particles}}
\centerline{\large\bf{using the Gravitational Wave Interferometers}}
\vspace{.15in}

\centerline{Adrian  Melissinos }

\centerline{\it{Department of Physics and Astronomy, University of Rochester }}

\centerline{\it{Rochester, NY 14627-0171, USA}}
\vspace{.1 in}
\centerline{10 March 2022}
\vspace{.10 in}
\begin{abstract}
\section{Abstract}
If ALPs exist they are expected to interact with the Electromagnetic field through a term of the form
$\mathcal{L}_{a\gamma\gamma} 
= -(g_{a\gamma\gamma}/4) F_{\mu\nu} \overset{\sim}{F_{\mu\nu}}\phi_a $.
 Therefore if light traverses a region where an ALP field exists, its refractive index will be modified. We propose to measure the refractive index of light and its angular dependence as the Earth rotates with respect to the direction of the ALP wind, as evidence for the presence of such a wind.
The LIGO interferometers are sensitive to differences in the refractive index of
the light ($\lambda = 1.064$\  $\mu$m) circulating in the two arms, and such differences were recorded 
during the S5 run, February 2006 to July 2007 \cite{Grossman,PLB}. They are due in part to the horizontal tidal gradients when they are aligned with one of the arms, imposing a shift on the 
frequency of the light circulating in that arm. In addition a very strong modulation
was observed at twice the Earth's orbital frequency, as can be seen in Fig.1. This can be understood  if the difference in refractive index between the two arms depends on the angle $\theta$ between the light propagation vector $\hat{k}$ and a ``wind" of ALP's incident on the Earth 
from a fixed direction \cite{axionwind}. A difference $\Delta (n_1 - n_2) \propto {\rm{sin^2}}(\theta_1) - {\rm{sin^2}}(\theta_2)$ reproduces the observed modulation.
We present the data as a function 
of the Earth's motion, discuss the magnitude of the observed refractive index,
$\Delta n = n -1 \approx 10^{-20}$, and conclude that such data can reveal 
the angular dependence and magnitude of the refractive index of the light circulating 
in the arms in the presence of an ALP ``wind". 

\end{abstract}

\section{Introduction}
The LIGO  interferometers are optimized for the detection of gravitational wave signals in the
range 40-1000 Hz. However the difference in the phase of the light returning
from the two arms to the detector (dark) port, that is the ``signal", can also be 
recorded over long time intervals by using detection at the ``free spectral range
frequency" as explained in detail in \cite{arXiv_2019}.
The signal measured at the detector port is the difference in the phase shift of the
light returning from the two arms \begin{equation} \Delta \phi = \delta\phi_1 
-\delta\phi_2 \end{equation} The individual phase shifts $\delta\phi_j$, j= 1,2 
can in principle arise from changes $\delta L_j$ in the effective path lengths
$L_j$ of the arms or a change in the  frequency of the light. The net phase shift
can also be affected by modifications $\delta\overline{n}_j$ in the effective refractive index 
$\overline{n}$ experienced by the light propagating in the two arms. The net 
phase shift on the $j_{th}$ arm for a single light traversal of length $2L$ can be 
expressed as
\begin{equation} \frac{\delta \phi^{(s)}_j}{2\pi} = \left(\frac{\delta L_j}{L} 
+ \frac{\delta f_c}{f_c} + \frac{\delta \overline{n}_j}{\overline{n}}\right)
\frac{2L}{\lambda} \end{equation} 
In the operating mode, the interferometer is ``locked" on a dark fringe by adjusting
the carrier frequency $f_c$ and the effective path lengths $L_j$ using feedback so that 
$\Delta\phi = 0$ is enforced at the detector port in the absence of a gravitational 
signal. Over a sufficiently large time interval $T$ compared to the time between 
successive feedback actions the integrated net phase change reduces to an integral
over changes in the difference \begin{equation} \Delta \overline{n}_{12} =
\delta\overline{n}_1 - \delta\overline{n}_2 \end{equation} of the effective 
refractive indices \begin{equation} \left<\frac{\Delta \phi}{2\pi}\right>=
\frac{1}{T} \int\limits_{t-T/2}^{t+T/2}\frac{\delta \phi_j}{2\pi}\quad   {\rm{or}}\quad
\frac{1}{T}\frac{2L}{\lambda}\int\limits_{t-T/2}^{t+T/2}\frac{\Delta\overline{n}}{n}dt
\end{equation} because the changes $\delta L_j$ and $\delta f_c$ are stochastic 
and average to zero when the interferometer is locked.

The above demonstrate that the interferometer does have sensitivity to time varying 
signals arising from a difference in the index of refraction of the light circulating
in each of the two orthogonal arms. The time variation of $\overline{n}_j$ can be
related to the Earth's daily (siderial) rotation angular frequecy, $\omega_{\oplus}
\approx 7.3\times 10^{-5}$ rad/s, its annual orbital angular frequency $\Omega_{\oplus}
\approx 2\times 10^{-7}$ rad/s, as well as to horizontal tidal gradients that
``red shift" the frequency of the light circulating in the arms \cite{arXiv_2019}.
Such signals involve frequencies many orders of magnitude below the optimized band.
At these low frequencies the instrumental noise prevents the extraction of the signal
by the conventional detection process. However by taking advantage of information 
circulating in the interferometer at a sideband frequency, and of the long integration 
time that is available, the low and very low frequency signals can be extracted. 
This is discussed in detail in \cite{arXiv_2014}.\\

We consider data taken by the H1 (Hanford, WA) interferometer during the S5 LIGO run 
over a period of 16 months from April 2006 to July 2007. These data wete reported in preliminary 
form in \cite{Grossman, PLB} and consist of a time series of the amplitude of the signal
at the dark port spaced at 64 second intervals. The signal amplitude is shown in
Fig.1 for the entire 16 months of data taking. There are periods of time when no data is available
because the interferometer was inoperative or had fallen out of lock. The signal amplitude
is directly proportional to
\begin{equation}  {\rm{Amplitude}}\propto\Delta n_{1,2} = \overline{n}_1 - \overline{n}_2 ,\end{equation}
namely the difference in the refractive index between the two arms.
 
Such a difference in refractive index can arise when a horizontal gravity gradient 
is present along one (but not the other) of the arms \cite{arXiv_2014}. Indeed the 
tidal forces have a horizontal component along the arms, typically \begin{equation}
g_{hor} \approx 10^{-7}g \approx 10^{-6}\ {\rm{ms^{-2}}} \end{equation} which is 
time dependent as the Earth and the Moon rotate. The frequency and amplitude of these 
components is well known \cite{Melchior}. When the horizontal tidal gradient is aligned
with one of the arms it imposes\footnote{This is a manifestation of the direct coupling 
of the gravitational gradient to the light circulating in the interferometer.}
a frequency shift (``red shift") on the light circulating in the arm. 
The resulting phase shift
with respect to the other arm, that is not aligned with the tidal gradient, is  given
for a single (round trip) traversal in the arm by \begin{equation}
\delta\phi^{(s)} = 2\int d\omega dt= 4\pi \nu_0\int \limits_0^L \frac{\delta \nu}{\nu} 
\frac{dx}{c} = \frac{4\pi}{\lambda_0}\int \limits_0^L \frac{\Phi}{c^2} dx = \frac{2\pi}
{\lambda_0} g_{hor} \frac{L^2}{c^2} \end{equation} 
where $\nu_0, \lambda_0$ refer to the carrier frequency, $L$ is the length of the arm and 
$\Phi = g_{hor}x$ is the gravitational potential. For $g_{hor}= 10^{-6}\ {\rm{m/s^2}}$ 
the phase shift imposed on the LIGO interferometer for a single traversal is

\begin{equation} \frac{\delta \phi^{(s)}}{2\pi} \approx 2\times10^{-10}  \end{equation} 

\section{The time-dependence of $\Delta n_{1,2}$}
To examine the time-dependence of the difference in refractive index between the two 
arms of the interferometer we spectrally (Fourier) analyze\footnote{Because the data is 
not continuous we must use the Lomb-Scargle algorithm \cite{Lomb-Scargle} which fits the data 
to a sine and cosine series.} the 14-month long time series shown in Fig.1. This reveals 
the presence of several discrete frequencies, centered around the Earth's daily and
twice daily rotation frequencies. The daily rotation frequencies are shown in Fig.2, and 
the twice daily in Fig.3. The frequency resolution is $\Delta f = 1/(4T_{total}) =
6\times 10^{-9}\ {\rm{Hz}}$ with\footnote{The factor of 4 is included because in the
spectral analysis the data was oversampled by that factor.} 
$T_{total} = 4.2\times 10^7\ {\rm{s}}$. The measured and known 
frequencies and amplitudes \cite{Melchior}
are in excellent agreement as also shown in Table I, with two exceptions: in the daily 
group (see Fig.2) the dominant line is at $f = 1.157\times 10^{-5}\ {\rm{Hz}}$, 
which corresponds to the exact daily 
(solar) rotation frequency, but the tidal line at this frequency, the S1 elliptic wave,{has an amplitude that is $0.01$ of the amplitude of the K1 declinational line\footnote{The K1 line is clearly resolved in Fig.2.},
at $f = 1.606 \times 10^{-5}\ {\rm{Hz}}$, namely it is completely unobservable.
 We conclude that the dominant line in the daily region is not understood and could be due
to human activity, or thermal effects.

The twice daily frequencies are shown in Fig.3 and the four lines are all attributed to
tidal gradients. They agree both in frequency (see Table I) as well as in magnitude with 
the known values \cite{Melchior}. The dominant line in this grouping is the Lunar principal wave
M2 with an amplitude of $91 \ \mu{\rm{gal}}, (1\ {\rm{gal} = 1\ {m/s}^2})$.
We calculate the horizontal component of the tidal force along the arms 
of the interferometer at the latitude of the Hanford site, and for the orientation of 
the two arms. For the dominant M2 line we find
$$F_{South} \approx 0.7\times 10^{-6}\ {\rm{ m/s^2}}\qquad \qquad \qquad
F_{West} \approx 10^{-6}\ {\rm{ m/s^2}}$$
Thus the phase shift induced by the M2 line for a single  traversal is 
$$\Delta \phi^{(s)}/2\pi = 1.2\times 10^{-10}$$
The measured power in the M2 line is obtained by integrating  the spectral line in Fig.(3) over frequency\footnote{Since the true line width is narrower
than the experimental resolution, the power is given by the area under the peak.} 
\begin{equation} \rm{M2 \qquad\qquad Measured\ power\ 3538\ counts}\qquad {\rm{ corresponds\ to}}
\quad \Delta \phi^{(s)}/2\pi = 1.2\times 10^{-10}\end{equation}
 Thus we can establish a relation between an observed power spectral density amplitude and the corresponding phase shift.

 \vspace{0.5in}
{\underline { Table I. \ \ Observed and known frequencies of the tidal
components (.Hz)}}

\begin{tabular}{c l|l|l} \hline

\qquad Symbol \qquad \qquad  & Measured \qquad \qquad & Predicted \qquad \qquad & Origin,  L=lunar; S=solar\\
\hline\\[0.2ex]
{\underline {Long period }}\\
 $\rm{Ss_a}$ & $6.536\times 10^{-8}$ & $6.338\times 10^{-8}$ & \rm{S declinational}\\[0.5ex]
 {\underline {Diurnal }}\\
 $\rm{O_1}$ & $1.07601\times 10^{-5}$ & $1.07585\times 10^{-5}$ & \rm{L principal lunar wave}\\
 $\rm{P_1}$ & $1.15384\times 10^{-5}$ & $1.15424\times 10^{-5}$ & \rm{S solar principal wave}\\
 $\rm{S_1}$ & $1.15741\times 10^{-5}$ & $1.15741\times 10^{-5}$ & \rm{S elliptic wave of} $\rm{^{s}K_1}$\\
 $\rm{^{m}K_1,^{s}K_1}$ & $1.16216\times 10^{-5}$ & $1.16058\times 10^{-5}$ & \rm{L,S declinational waves}\\
 [0.5ex]

{\underline {Twice-daily }}\\
 $\rm{N_2}$ & $2.19240\times 10^{-5}$ & $2.19442\times 10^{-5}$ & \rm{L major elliptic wave of $\rm{M_2}$}\\
 $\rm{M_2}$ & $2.23639\times 10^{-5}$ & $2.23643\times 10^{-5}$ & \rm{L principal wave}\\
 $\rm{S_2}$ & $2.31482\times 10^{-5}$ & $2.31481\times 10^{-5}$ & \rm{S principal wave}\\
 $\rm{^{m}K_2,^{s}K_2}$ & $2.31957\times 10^{-5}$ & $2.32115\times 10^{-5}$ & \rm{L,S declinational waves}\\

\end{tabular}\\
\vspace{0.3 in}
 
As can be seen by inspection of Fig.1, superimposed on the daily and twice daily oscillations
is a modulation at much lower frequency. The spectral analysis reveals that this modulation
is at twice the Earth's orbital frequency, within the frequency resolution as shown in Fig.4. 
The observed frequency is 
\begin{equation} f_{observed} = (6.239 \pm 0.6)\times 10^{-8} \ {\rm{Hz}}\quad {\rm{as\ 
compared\ to }}\quad 2f_{\oplus{\rm{orbital}}} =6.338\times 10^{-8}\ {\rm{Hz}} \quad\quad    
  \end{equation}

We attribute the observed time dependence of the difference in the refractive indices, $\Delta n$ to the interaction of the light in the interferometer arms with a ``wind" of ALPs that has a fixed
direction in space. The wind arises because of the rotation of the spiral arm of the Galaxy
  that contains the Earth, through the static ALP background that permeates the Galaxy 
	\cite{Turner, SAO}.
We seek a dispersion relation that can reproduce the data. When light propagates through a
cold ALP background, the refractive index is modified as shown by Espriou and Cerillo \cite{
Espriou} but has no angular dependence. 


Recently McDonald and Ventura have published a  dispersion relation for
light propagating through an ALP wind \cite{McDonald_1, McDonald_2}. 
Here $(p_0,\vec{p})$ is the 4-vector of the ALP, and $\theta$ the angle between the ALP
momentum and the direction of the light, $\hat{k}$. Since the angle $\theta$ changes 
continuously due to the Earth's rotation with respect to the direction of the wind,
we expect a corresponding change in the refractive indices along the two arms of the 
interferometer. The refractive
index is given by  Eq.(25) in \cite{McDonald_1}, \begin{equation}
\frac{|k| -\omega}{|k|} =-\delta n = - \frac{g^2_{a\gamma\gamma}}
{16|k|^2} 
\left[\dot{a}^2 + (\hat{k}\cdot\nabla{a})^2 - 2|\nabla{a}|^2\right]\end{equation}
\begin{equation}\delta n =\frac{g^2_{a\gamma\gamma}}{16|k|^2} \left[\omega_a^2 a_0^2
(1+\beta_a^2 {\rm{cos}}^2(\theta)-2\beta_a^2\right]\end{equation} 
\begin{equation}\delta n =\frac{g^2_{a\gamma\gamma}}{16|k|^2} 2\rho_a \left[
1-\beta_a^2 -\beta_a^2({\rm{sin}}^2(\theta))\right]\end{equation}
In the above, the authors introduce the ALP-photon-photon coupling $g_{a\gamma\gamma}$, 
and state that the refractive index arises from the quadratic term in the dispersion
relation; this is supported  by referring to Fig.(2)  of \cite{McDonald_2}.
Further, $a_0$ is the amplitude of the ALP field, $\beta_a$ the ALP velocity
and $\omega_a\sim m_a$ the ALP energy, approximately equal to the ALP mass $m_a$,
$a_0^2 m_a^2 =\rho_a$ is the ALP energy density which is well known from 
observational data to be $\rho_a = 10^{-42}\ {\rm{GeV^4}}$ \cite{ALP_Density}.\\

We have calculated the angles $\theta_{1,2}$ of the two arms of the Hanford interferometer,
properly oriented, and
accounting for the Earth's rotation, orbital motion and the motion of the solar system
in the direction right ascension $\alpha = 85^{\circ}$, declination $\delta =-29^{\circ}
$ with a velocity
in the SCCEF frame $V_{\odot}$ = 500 km/sec. 

It follows from Eq.(13) that \begin{equation} \Delta_{n1,n2} = \frac{g^2_{a\gamma\gamma}}
{16|k|^2} 2\rho_a [{\rm{sin}}^2(\theta_1)-{\rm{sin}}^2(\theta_2)] \end{equation}
Using the calculated
angular dependence of the orientation of the interferometer arms, we have modeled
Eq.(14). Spectral analysis of the modeled time series reveals a  strong line at 
$f = 2\Omega_{\oplus}$, twice the Earth's orbital rotation frequency, as
observed in the data, confirming the angular dependence of the refractive index
predicted by Eq.(14). An arbitrarily normalized fit of Eq.(14) to the data, as shown in Fig.(5),
is  satisfactory.

\section{The magnitude of the refractive index.}

From the power spectral density for the dominant line at the twice yearly orbital frequency, Fig.(4), and using the conversion 
factor of Eq.(9) we find that the phase shift $\Delta\phi^{(s)}$
for light at $\lambda \sim 1\ \mu m$ 
propagating through the ALP wind incident on the Earth is of order
\begin{equation} \Delta\phi^{(s)} = 8\times 10^{-10\pm 1}\end{equation}
The refractive index is simply related to the phase shift
$$ \Delta n= n-1 =\frac{\lambda}{2L} \Delta \phi^{(s)}$$
and therefore we estimate that
 \begin{equation} \Delta n = n-1 =10^{-20\pm 1}\end{equation} 
We expect the refractive index to depend on the ALP-photon-photon 
coupling, the ALP mass,  the known ALP mass density, and the velocity of the
ALP wind which we can take as $\beta_a = 10^{-3}$.


 The magnitude of 
the angle dependent part is\begin{equation} \Delta n(\theta) = \frac{g_{a\gamma\gamma}^2}
{16|k|^2}2\rho_a\beta_a^2\end{equation} If we use the current upper limit \cite{CAST}
$g_{a\gamma \gamma} < 10^{-10} {\rm{GeV^{-1}}}$, $k =\ 1\ {\rm{eV}}$, and $\rho_a =10^{-42}
\ {\rm{GeV}}^4$ and $\beta_a = 10^{-3}$ we find $\Delta n(\theta) = 10^{-51}$, as compared
to the measured value $\Delta n(\theta) = 10^{-20}.$ Clearly Eq.(17) must be modified to 
describe the data. In the absence of theoretical guidance we use a dimensional arguement
and divide Eq.(17) by $(m_a/\omega)^2$;
in this case the observed refractive index can be recovered for
  \begin{equation}
\frac{g_{a\gamma \gamma}/ {\rm{GeV^{-1}}}}{m_a/{\rm{eV}}} = 3\times 10^{5} \end{equation}
which for $g_{a\gamma \gamma} = 10^{-10}\ {\rm{GeV^{-1}}}$ \cite{CAST} leads to $m_a = 3\times 10^{-16}
\ {\rm{eV}}$, which is not experimentally excluded but significantly different from
the range predicted by the theoretical models \cite{PDG}.


We conclude that the gravitational Interferometers can be used to measure the 
refractive index of the light circulating in the arms to a precision of order 
$$\Delta n = n-1 \approx 10^{-20}$$ and its angular dependence. Such data can
 establish the presence of an ALP wind incident on the Earth, and the direction of 
the wind.\\

\newpage
{\large{\bf{Acknowledgments:}}}
I am indebted to the LIGO team that made these measurements possible and in particular 
to D. Sigg, F.J. Raab, W.E. Butler, C. Forrest, T. Fricke and S. Giampanis who were 
involved in the the design, installation and operation  of the fsr channel  and in the
analysis of the data. I also thank A. Kostelecky for insights on the effective refractive
index in interferometers, and G. Ruoso and E. Milotti for constructive comments on a 
previous version of this note.\\

\vspace{-3 in} 
\begin{figure}[H]
\centering

\hspace{-2 in}
\includegraphics[width=130mm,height=80mm]{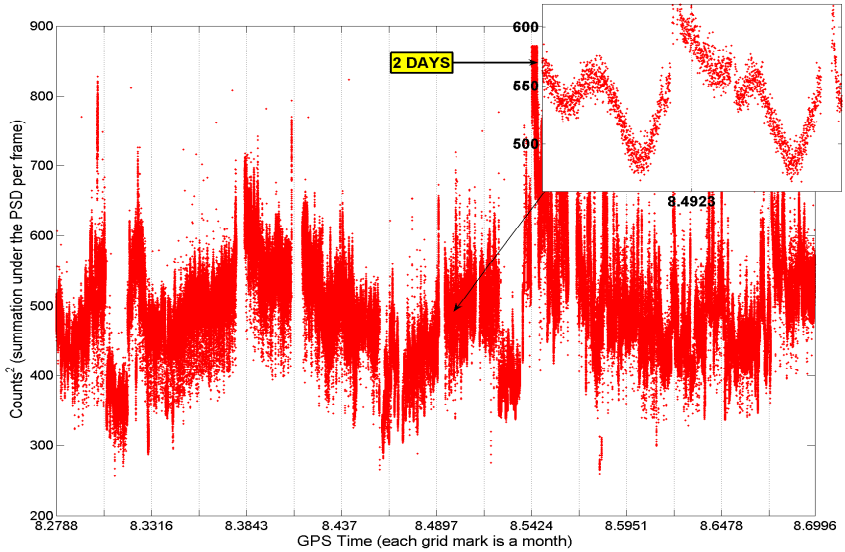}
\vspace{0.25in}
\caption{The integrated power spectral density in +/-
200 Hz of the fsr as a function of time for 16 months
during the S5 run. The daily and twice daily modulation
can be seen in the inset. Vertical lines are at monthly intervals.}
\end{figure}

\begin{figure}[H]
\centering
\includegraphics[width=130mm,height=60mm]{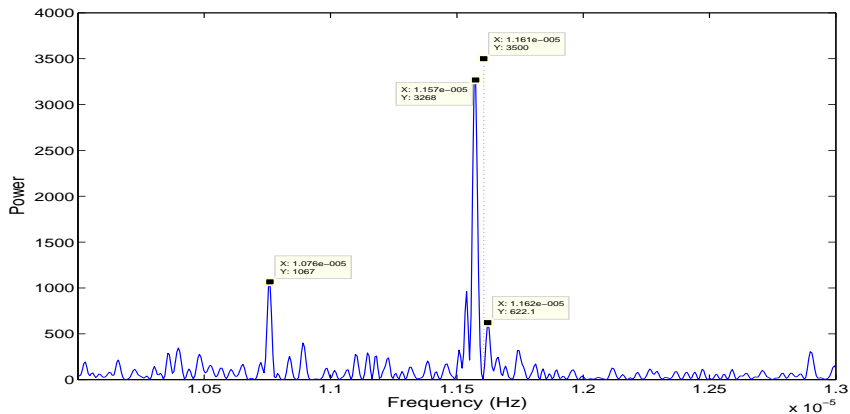}
\vspace{0.6 in}
\caption{Power spectrum in the daily frequency region. The dominant line 
at a frequency \newline f = 1.157e-5 is at exactly at the daily solar 
rotation frequency, and is not understood at this time, but must be due to the 
interaction of the light in the 
arms with an external agent as the Earth rotates.The siderial frequency is
indicated by the dashed red line.}
\end{figure}

\begin{figure}[H]
\centering
\vspace{-0.5 in}
\hspace{-1.5 in}
\includegraphics[width=130mm,height=60mm]{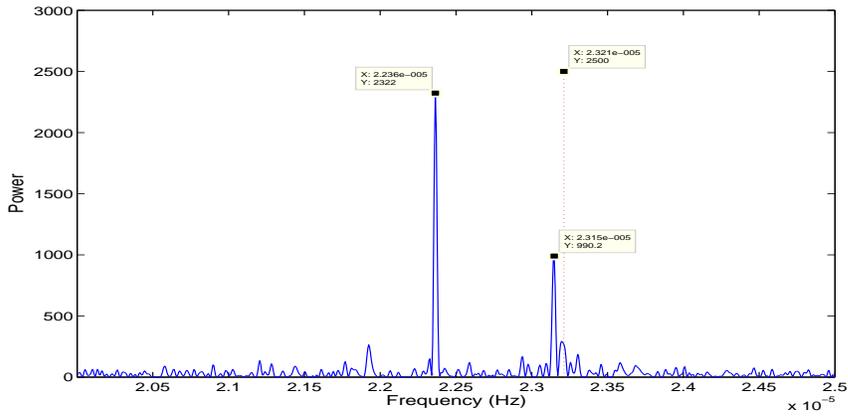}
\vspace{0.6 in}
\caption{Power spectrum in the twice daily frequency region. All lines can
be attributed to tidal gradients; see Table 1.}
\end{figure}

\begin{figure}[H]
\centering
\includegraphics[width=130mm,height=60mm]{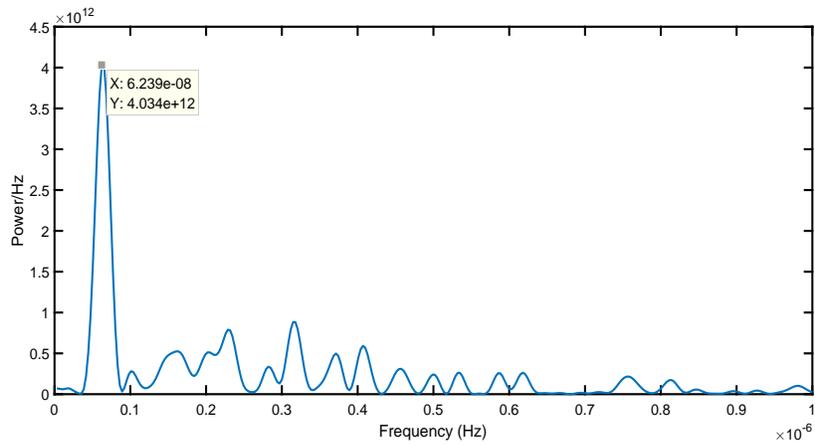}
\caption{Power spectrum of the integrated power spectral density 
in the very low frequency region. The spectral line is at  
 twice the Earth's yearly orbital frequency within the measurement error.}
\end{figure}

\begin{figure}[H]
\centering
\hspace{-1 in}
\includegraphics[width=130mm,height=80mm]{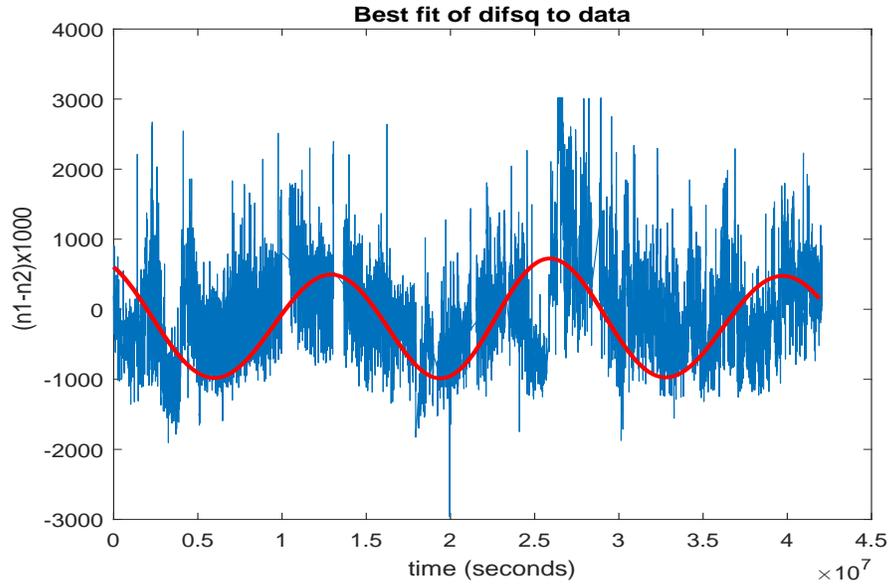}
\caption{Fit of the predicted envelope of the daily oscillations (red)
$[{\rm{sin}}^2(\theta_1) - {\rm{sin}}^2(\theta_2)]$ to the data. Time in seconds starts at winter equinox of 2006. Normalization is arbitrary.}
\end{figure}

\end{document}